\documentstyle[12pt,psfig]{article}   

\newcommand{\beq}{\begin{equation}}
\newcommand{\eeq}{\end{equation}}
\newcommand{\bea}{\begin{eqnarray}}
\newcommand{\eea}{\end{eqnarray}}
\newcommand{\smallw}{{\scriptscriptstyle W}} %

\newcommand{\mw}{M_\smallw}
\newcommand{\muw}{\mu_\smallw}

\newcommand{\Gop}[1]{Q_{#1}}
\def\f{\frac}
\def\al{\alpha_s}
\def\bsg{BR_\gamma}
\def\bsl{BR_{SL}}
\def \mt   {m_t}
\def \mc   {m_c}
\def \Cp   {C_{7_{OS}}}
\def \Cg   {C_{8_{OS}}}

\def \ms   {m_s}
\def \mb   {m_b}

\def\slash#1{\setbox0=\hbox{$#1$}#1\hskip-\wd0\dimen0=5pt\advance
       \dimen0 by-\ht0\advance\dimen0 by\dp0\lower0.5\dimen0\hbox
         to\wd0{\hss\sl/\/\hss}}
\def\gequiv{\raise 0.4ex \hbox{$>$} \kern -0.7 em \lower 0.62 ex \hbox{$\sim$}}
\def\gappeq{\mathrel{\rlap {\raise.5ex\hbox{$>$}}
{\lower.5ex\hbox{$\sim$}}}}

\def\pl#1#2#3{{\it Phys. Lett. }{\bf B#1~}(19#2)~#3}
\def\zp#1#2#3{{\it Z. Phys. }{\bf C#1~}(19#2)~#3}
\def\prl#1#2#3{{\it Phys. Rev. Lett. }{\bf #1~}(19#2)~#3}

\def\pr#1#2#3{{\it Phys. Rev. }{\bf D#1~}(19#2)~#3}
\def\np#1#2#3{{\it Nucl. Phys. }{\bf B#1~}(19#2)~#3}
\def\mpl#1#2#3{{\it Mod. Phys. Lett. }{\bf #1~}(19#2)~#3}

\def\ptp#1#2#3{{\it Prog. Theor. Phys. }{\bf #1~}(19#2)~#3}

\begin{document}              

\begin{titlepage}
\begin{flushright}
        \small
        CERN-TH/97-279\\
        MPI-PhT/97-64\\
        hep-ph/9710335  \\
        October 1997
\end{flushright}

\begin{center}
\vspace{1cm}
{\large\bf Next-to-Leading QCD Corrections to $B\to X_s \gamma$:\\
Standard Model and Two-Higgs Doublet Model}

\vspace{0.5cm}
\renewcommand{\thefootnote}{\fnsymbol{footnote}}
{\bf    M.~Ciuchini$^a$,
        G.~Degrassi$^b$\footnote
                {Permanent Address: Dipartimento di Fisica, Universit{\`a}
                  di Padova, Padova, Italy.},
        P.~Gambino$^c$\footnote
                {Present address: Technische Universit{\"a}t M{\"u}nchen,
                Physik Dept., D-85748 Garching, Germany.},
        G.F.~Giudice$^b$\footnote
                {On leave of absence from INFN, Sez. di Padova, Italy.}}
\setcounter{footnote}{0}
\vspace{.8cm}

{\it
        $^a$ INFN, Sezione Sanit\`a, V.le Regina Elena, 299, I-00186 Roma,
        Italy\\
\vspace{2mm}
        $^b$ Theory Division, CERN CH-1211 Geneva 23, Switzerland \\
\vspace{2mm}
        $^c$ Max Planck Institut f{\"u}r Physik, W. Heisenberg Institut,\\
        F{\"o}hringer Ring 6, M{\"u}nchen, D-80805, Germany }
\vspace{1cm}

{\large\bf Abstract}
\end{center}
We present the QCD corrections to the matching conditions of the $\Delta
B =1$ magnetic and chromo-magnetic operators in the Standard Model and
in two-Higgs doublet models. We use an off-shell matching procedure
which allows us
to perform the computation using Taylor series in the external momenta,
instead of asymptotic expansions.
In the Standard Model case, we confirm previous results derived
on-shell and we obtain $BR( B\to X_s \gamma )=(3.62\pm 0.33)
\times 10^{-4}$. In the case of
the usual two-Higgs doublet model, we show that going from the leading
to the next-to-leading order result improves the CLEO bound
on the charged-Higgs mass from 260 GeV to 380 GeV.
This
limit is very sensitive to the definition of errors and we carefully
discuss the theoretical uncertainties. Finally, in the case of the two-Higgs
doublet model in which both up- and down-type quarks couple to the same
Higgs field, the theoretical prediction for $BR( B\to X_s \gamma )$ can
be reduced by at most 20\% with respect to the Standard Model value.

\noindent


\end{titlepage}
\newpage
\eject
\section{Introduction}
\label{sec:introduction}
Interest in radiative $B$ decays has been renewed in recent years,
after the first measurements of the exclusive $BR(B\to 
K^\star\gamma)$~\cite{cleoks} and
the inclusive $BR(B\to X_s\gamma)$~\cite{cleo} were performed
by the CLEO
collaboration.
Theoretically these processes are
interesting because they are sensitive to physics beyond the Fermi scale
already at the leading order (LO) so that they are potentially well suited
to probe new physics.  
Moreover these are ``rare'' processes with quite large 
branching ratios, of the order of  $10^{-4}$, partly because they are
strongly enhanced by the QCD corrections.
Finally, the non-perturbative corrections to the inclusive
decay turn out
to be small~\cite{sav,tran,vol,isi,mcc}, thus lending greater
confidence
on the perturbative calculation.

Experimentally, the measurement at the $\Upsilon(4S)$~\cite{cleo}
\beq
BR(B\to X_s\gamma)= (2.32\pm 0.57\pm 0.35)\times 10^{-4}
\eeq
was found in good agreement with the LO predictions~\cite{buras1,ciuchini1},
which however suffer from a large theoretical uncertainty, $\sim 30-40\%$,
coming mainly from the choices of the renormalization and matching scales,
which are not well defined at the LO.

Recently the next-to-leading order (NLO) calculation in the Standard Model
(SM) has been completed~\cite{yao,ali,wyler,misiak1}, raising the
accuracy to the $10\%$ level. The new predictions~\cite{misiak1,buras2} 
turn out to be
slightly larger than the CLEO measurement.
However preliminary ALEPH results~\cite{aleph} now
indicate a larger branching ratio measured at the $Z^0$
 resonance,
\beq
BR(B\to X_s\gamma)=(3.38\pm 0.74\pm0.85)\times 10^{-4}~.
\eeq
Since this result is only preliminary and a complete analysis is now in 
progress, we will not use it in our discussion. However, if this result
is confirmed, it shows that the experimental situation is not settled.

In this paper, we present a new computation of the matching conditions
of the $\Delta B=1$ magnetic and chromo-magnetic operators in the SM and
in the two-Higgs-doublet model at the NLO, using an off-shell
procedure. We confirm the previous
results
in the SM~\cite{yao,greub,buras2}, obtained by on-shell matching, 
and present the charged Higgs contribution
to the Wilson coefficients in the two-Higgs-doublet model.
The paper is organised as
follows. In sec.~\ref{sec:matching}, we give some general formulae useful
for implementing the matching conditions and discuss the peculiarities of our
approach. Section ~\ref{sec:outline} is devoted to some technical details 
of the calculation. In sec.~\ref{sec:SM}, the relevant formulae for
computing the NLO corrections to 
$BR(B\to X_s\gamma)$ are collected and a new
phenomenological analysis in the context of the
SM is presented. Sec.~\ref{sec:2HDM} contains the
new coefficients computed in the two-Higgs doublet model
and a discussion of the
bounds enforced by the measurement of $BR(B\to X_s\gamma)$ on the parameter
space $(M_{H^+}, \tan\beta)$ of the model.
Particular care is devoted to the definition of the theoretical errors,
which play a major r\^ole in the phenomenological analysis, especially
in the two-Higgs doublet model.
Finally, in sec.~\ref{sec:conclusions}, we summarise our
results and conclusions.

\section{The Matching Conditions at the NLO}
\label{sec:matching}
In this section we  discuss the determination of the matching
condition of the $\Delta B=1$ effective Hamiltonian $H_{eff}$ at the
NLO.

The $\Delta B=1$ effective Hamiltonian is given by
\beq
H_{eff}=-\frac{4G_F}{\sqrt{2}} V^\star_{ts} V_{tb} \vec Q(\mu)\cdot
\vec C(\mu)~,
\label{eq:heff}
\eeq
where $G_F$ is the Fermi coupling constant and $V$ is the 
Cabibbo-Kobayashi-Maskawa (CKM)
quark mixing matrix.
$\vec Q(\mu)$ is the row vector containing the renormalised
dimension-six local operators which form the basis for the operator product
expansion (OPE) of the $\Delta B=1$ weak-current product and $\vec
C(\mu)$ are the
corresponding Wilson coefficients.
The determination of the effective Hamiltionan requires three steps:
i) choice of the operator basis;
ii) computation of the initial conditions for the Wilson coefficients by
matching suitable amplitudes computed in the ``full'' and in the effective
theory renormalised at a matching scale $\muw$;
iii) calculation of the relevant
anomalous dimension matrix which determines the dependence of the Wilson
coefficients on the renormalization scale $\mu$ through the renormalization
group equations.

In the following we present the computation of the matching conditions for the
(chromo-) magnetic
operators in the case of the SM and of the two-Higgs-doublet
model. The calculation in the SM, originally carried out
in ref.~\cite{yao}, has been already recently confirmed~\cite{greub,buras2}.
The case of the two-Higgs doublet model is a novelty. 

A remarkable difference between our calculations and previous ones is 
that we work with the off-shell effective
Hamiltonian, matching off-shell matrix elements, while the authors
of refs.~\cite{yao,greub,buras2}
always use on-shell conditions. The advantage of an off-shell matching
relies on the fact that one can choose a suitable kinematical configuration
such that the relevant 
Feynman diagrams can be evaluated using ordinary Taylor expansions in the
external momenta, in contrast to the  on-shell case~\cite{greub}, in
which asymptotic expansions~\cite{smirnov} must be employed.
However, this requires to work with an enlarged operator basis. 

The off-shell operator basis relevant for our calculation is~\cite{misiak1,
curci,simma}:
\beq
\begin{array}{rl}
Q_1 = & (\bar{s}_L\gamma_{\mu}t^a c_L) (\bar{c}_L \gamma^{\mu} t^a b_L) 
\vspace{0.2cm} \\
Q_2 = & (\bar{s}_L\gamma_{\mu} c_L) (\bar{c}_L\gamma^{\mu} b_L) 
\vspace{0.2cm} \\
Q_3 = & (\bar{s}_L\gamma_{\mu} b_L) \sum_q (\bar{q}\gamma^{\mu} q)     
\vspace{0.2cm} \\
Q_4 = & (\bar{s}_L \gamma_{\mu} t^a b_L) \sum_q (\bar{q}\gamma^{\mu} t^a q)     
\vspace{0.2cm} \\
Q_5 = & (\bar{s}_L\gamma_{\mu}\gamma_{\nu}\gamma_{\rho} b_L)
        \sum_q (\bar{q}\gamma^{\mu}\gamma^{\nu}\gamma^{\rho} q)
\vspace{0.2cm} \\
Q_6 = & (\bar{s}_L\gamma_{\mu}\gamma_{\nu}\gamma_{\rho} t^a b_L)
        \sum_q (\bar{q}\gamma^{\mu}\gamma^{\nu}\gamma^{\rho} t^a q)
\vspace{0.2cm} \\
Q_7  = &  \frac{e}{16\pi^2} m_b (\bar{s}_L \sigma^{\mu \nu} b_R) F_{\mu \nu} 
\vspace{0.2cm} \\
Q_8  = &  \frac{g_s}{16\pi^2} m_b (\bar{s}_L \sigma^{\mu \nu} t^a b_R)
 G_{\mu \nu}^a 
\vspace{0.2cm} \\
Q_9 = & \frac{i}{16\pi^2}\bar{s}_L \not\!\! D\not\!\! D\not\!\! D b_L
\vspace{0.2cm} \\
Q_{10} = & \frac{i e}{16\pi^2}\bar{s}_L \left\{\not\!\! D, \sigma_{\mu\nu}
F^{\mu\nu}\right\} b_L 
\vspace{0.2cm} \\
Q_{11} = & \frac{e}{16\pi^2}\bar{s}_L \gamma_\nu b_L D_\mu F^{\mu\nu}
\vspace{0.2cm} \\
Q_{12} = & \frac{1}{16\pi^2}\mb \,\bar{s}_L \not\!\! D\not\!\! D b_R 
\vspace{0.2cm} \\
Q_{13} = & \frac{i g_s}{16\pi^2}\bar{s}_L \left\{\not\!\! D, \sigma^{\mu\nu}
t^a G^a_{\mu\nu}\right\} b_L  
\vspace{0.2cm} \\
Q_{14} = & \frac{g_s}{16\pi^2}\bar{s}_L \gamma^\nu t^a b_L D^\mu
G^a_{\mu\nu}~,
\vspace{0.2cm}
\end{array}
\label{eq:basis}
\eeq
where $g_s$ ($e$) is the strong (electromagnetic) coupling constant,
$q_{L,R}$ are the chiral quark fields, $G^a_{\mu\nu}$ ($F_{\mu\nu}$) is
the QCD (QED) field strength, $D_\mu$ is the $SU(3)_C\times U(1)_Q$ covariant
derivative and $t^a$ are the colour matrices, normalised such that
$\mbox{Tr}\left(t^at^b\right)=\delta^{ab}/2$. On-shell $Q_9,...,Q_{14}$ 
are no longer independent operators and, by applying the
equations of motion (EOM), they
can be written as
linear combinations of the operators $Q_1,...,Q_8$.

In order to calculate the Wilson coefficients of
the magnetic operators, one
has to match a suitable amplitude computed both in the ``full'' and the
effective theory between the same external states at a matching scale
$\muw$. Schematically, the condition is
\beq
\langle s\gamma \vert J^\mu J^\dagger_\mu\vert b\rangle=\langle s\gamma\vert
\vec Q(\muw) \vec C(\muw)\vert b\rangle~,
\label{eq:matching}
\eeq
where $J^\mu J_\mu^\dagger$ denotes the weak current product which is
responsible for the $\Delta B=1$ transitions in the ``full'' theory.

At the NLO, retaining only the leading terms in $1/M_W^2$,
the relevant amplitude computed in the ``full'' theory is given by
\beq
\langle s\gamma \vert J^\mu J_\mu^\dagger\vert b\rangle=\langle s\gamma \vert
\vec Q^{(0)}\vert b\rangle\cdot (\vec T^{(0)}+
\frac{\alpha_s}{4\pi} \vec T^{(1)})~,
\eeq
where $\langle \vec Q^{(0)}\rangle$ are the tree-level matrix elements of
the operators in eq.~(\ref{eq:basis}).
Introducing the perturbative expansion
\beq
\vec C(\mu)=\vec C^{(0)}(\mu)+\frac{\alpha_s}{4\pi} \vec C^{(1)}(\mu)
\eeq
and defining a matrix $r$ such that
\beq
\langle s\gamma \vert\vec Q(\muw)\vert b\rangle=\left( r^{(0)}+
\frac{\alpha_s}{4\pi} r^{(1)}\right) \langle s\gamma \vert\vec Q^{(0)}
\vert b\rangle~,
\eeq
the matching condition, eq.~(\ref{eq:matching}), becomes
\beq
\vec T^{(0)}+\frac{\alpha_s}{4\pi} \vec T^{(1)}= r^{(0)T}\vec C^{(0)}+
\frac{\alpha_s}{4\pi}( r^{(1)T}\vec C^{(0)}+ r^{(0)T}\vec C^{(1)})~,
\eeq
namely
\beq
\vec C^{(0)}=( r^{(0)T})^{-1}\vec T^{(0)}~,\qquad
\vec C^{(1)}=( r^{(0)T})^{-1}(\vec T^{(1)}- r^{(1)T}
( r^{(0)T})^{-1}\vec T^{(0)})~.
\label{eq:coeff}
\eeq
Notice that i) in the presence of the magnetic operators, $r^{(0)}$
is not the identity matrix. Indeed the operator $Q_2$ has non-vanishing
matrix element on $Q_{11}$ and $Q_{14}$ at the LO, through a one-loop
diagram. Therefore the elements $ r^{(0)}_{2,11}$ and
$ r^{(0)}_{2,14}$ do not vanish. ii) $\vec T^{(1)}$ contains terms
proportional to $\log(m^2/M_W^2)$ and similarly $ r^{(1)}$ contains
$\log(m^2/\muw^2)$, $m^2$ being the infrared regulator, so that the matching
conditions depend on $\log(\muw^2/M_W^2)$, but are independent of the
infrared regulator. These $\log$ terms compensate a change of the
matching scale $\muw$ at the NLO, as stressed in ref.~\cite{buras2}.
Since the OPE must be valid on any matrix element, the Wilson coefficients
cannot depend on the external states used to impose the matching
conditions~\cite{wilson,witten}. For this reason our choice of matching
on off-shell states is completely justified.

According to eq.~(\ref{eq:coeff}), we first have to perform the tree-level
matching by computing the relevant penguin diagrams 
for the various operator insertions. Only $C_2$ and $C_7$--$C_{14}$ are
found to be non-vanishing at the LO. The LO
matching of the off-shell effective Hamiltionian, $(r^{(0)T})^{-1}
\vec T^{(0)}$, has already been computed in ref.~\cite{curci}.
We have repeated the calculation and confirmed the results in eq.~(27) of
ref.~\cite{curci}.

Then we can proceed to calculate the
finite parts of the two-loop diagrams, obtained by attaching gluons
to the LO diagrams.
Since we are
only interested in on-shell magnetic operators, at the NLO we just need
the coefficients of the operators which have a projection on the magnetic
ones. Using the EOM, one finds
\bea
 \Cp &=& C_7+C_{10}\nonumber\\
 \Cg &=& C_8+C_{13}~,
\eea
$\Cp$ and $\Cg$ being the coefficients of the on-shell
magnetic and chromo-magnetic dipole operators respectively.
Given the texture of $r^{(0)}$, we have to compute only 
$\vec T^{(1)}_i$, and $r^{(1)}_{ij}$ with $i=7,8,10,13$ and
$j=1,\dots,14$.

In the on-shell basis, it is more convenient to use the ``effective''
(chromo-) magnetic coefficients, defined as~\cite{ciuchini1,buras1,misiak1}
\bea
\Cp^{eff}=\Cp + \sum_{i=1}^6 y_i C_{i_{OS}} \nonumber\\
\Cg^{eff}=\Cg + \sum_{i=1}^6 z_i C_{i_{OS}}~, 
\eea
where $y$ and $z$ are scheme-dependent constant vectors, given by
\bea
y_i^{NDR}=(0, 0, -\frac{1}{3}, -\frac{4}{9}, -\frac{20}{3}, -\frac{80}{9})
\nonumber\\
z_i^{NDR}=(0, 0, 1, -\frac{1}{6}, 20, -\frac{10}{3}),
\eea
in NDR. These definitions guarantee that the LO matrix element of the
effective Hamiltonian for the transition $b\to s\gamma$ ($b\to s g$) is
proportional to $\Cp^{eff}$ ($\Cg^{eff}$) only.

Before outlining in the following section some technical details of
the calculation, we discuss here some general features.

\begin{enumerate}
\item We have regularized the ultraviolet divergences using 
naive dimensional regularization (NDR) with anti-commuting $\gamma_5$ and
defined the renormalized operators in the $\overline{\mbox{MS}}$ scheme.

\item We use a background-field gauge~\cite{abbott} for QCD
which avoids the appearance of non-gauge-invariant operators in the off-shell
basis~\cite{collins}. Moreover, in this gauge, the external-gluon-field
renormalization cancels out against the renormalization of the 
strong-coupling constant which is present in the definition of the operators 
with an external gluon field. In the electroweak sector we employ
the Fujikawa gauge~\cite{fuj,deshpande}.
\item Flavour non-diagonal quark self-energies $\Sigma_{ij}(p)$
appear at one loop in the
``full'' theory. These can produce dimension-four operators in the effective
theory, see {\it e.g.}~ref.~\cite{simma}. However it is well
known~\cite{altarelli} that the appearance of dimension-four operators can
be avoided by choosing suitable renormalization conditions in the ``full''
theory, which cancel the momentum-independent part of the off-diagonal
self-energies. This can be done by imposing that, for on-shell quark states
$q(p)$, $\bar q_i(p)\Sigma_{ij}(p)=\Sigma_{ij}(p)q_j(p)=0$, see
{\it e.g.} ref.~\cite{deshpande}.
\item Light-quark masses play the r\^ole of infrared regulators.
However  Feynman diagrams involving the
three-gluon vertex present additional infrared divergences that are 
not regulated by this procedure. The simplest
regulators in our approach are masses, thus we have computed these diagrams
with a fictitious gluon mass. 
Since the gluon mass is bound to violate the QCD Ward-Takahashi identities at
some order of the perturbative expansion~\cite{collins}, we have repeated the
calculation regularising these infrared divergences using the dimensional
regularization. We have found that, in this particular calculation, the gluon
mass is a ``well-behaved'' infrared regulator.

\item A final observation concerns the matrix elements of $Q_3$ and $Q_4$.
These operators have a scheme-dependent ${\cal O}(\alpha_s^0)$ matrix element
between $b$ and $s\gamma$, which is non-vanishing in
NDR~\cite{misiak2,ciuchini2}. 
Their contribution has to be explicitly considered in the on-shell
matching procedure~\cite{yao,greub,buras2}. This is however not necessarily
the case for off-shell matching conditions,
even if the same operators are also present in the off-shell basis. Indeed, 
these operators have vanishing off-shell initial conditions up to the NLO. The
contribution coming from $Q_{3,4}$ to the on-shell matching is provided
off-shell by the ${\cal O}(\alpha_s)$ matrix element of $Q_{11,14}$. The EOM
account for the mismatch in the power of $\alpha_s$.
\end{enumerate}

\section{Outline of the Calculation}
\label{sec:outline}

We work in an off-shell kinematical configuration
in which all external momenta are assumed to be much smaller than the 
masses of the particles running inside the loops. However, in this framework
$\ms$ is kept only when it is needed as an infrared regulator. 

To extract 
$\Cp$ (and analogously $\Cg$) 
we notice that the most general vertex diagram having as  external 
particles two fermions and one photon can be decomposed in $12 \times 2 $
objects, where 12 are the possible combinations of
one Lorentz index and two momenta (like, for example, $\slash{p} \, q^\mu$
with $p$ and $q$ independent momenta)  while  2 takes into account
the two possible chiralities. Using  standard techniques~\cite{gm2}
one can construct projectors for each of these 24 objects and then take the
appropriate combination of them which, after the use of EOM and the Gordon 
identity, contributes to the on-shell magnetic  operator. 
We observe that these projectors  depend on the kinematical
configuration assumed. We find it convenient to work in a ``scaled'' 
configuration defined by $ q^2 =0, \: p \cdot q = p^2/2$
where $p$ and $q$ are the momenta carried by the $b$ quark and
the photon respectively, under the assumption 
$p^2 \ll \ms^2, \, \mb^2$. In this configuration the 
projector for the tensor part of the on-shell (chromo-)magnetic operator is
\bea
P_{(7,8)}^\mu & = &  \frac1{2\,{p^4}\,(n -2 ) }
\left(
{p^2}\,\gamma^\mu \slash{p} - {p^2}\,\gamma^\mu \, \slash{q} - 
      \mb\,\gamma^\mu \slash{q}  \slash{p} \, + 2 \slash{p} \slash{q}\, 
      p^\mu -  2\,{p^2}\,p^\mu \right. \nonumber\\
&& ~~~~~~~~~~~~~~~ +     \left. \mb\,n\,\slash{p} \,q^\mu - 
   \mb\,n\,\slash{q}\,q^\mu + 
      n\, \slash{q} \slash{p} \,q^\mu \right) \label{proje}
\eea
and works by contracting it with the amplitudes and taking the trace.
In eq.~(\ref{proje}) $n$ is the dimension of the space-time and $\mu$ 
is the index carried by the photon. As it can be seen from
eq.~(\ref{proje}) the projector 
contains powers of $p^2$ in the denominator. However, because of our 
assumption $p^2 \ll \ms^2, \, \mb^2$,
we can perform  ordinary Taylor series to eliminate
the unphysical poles when $ p^2 \rightarrow 0$. Consequently, the actual 
computation of the graphs both for the
``full" and the effective theory is reduced to the calculation of massive 
self-energy integrals with no external momenta. Explicit results on two-loop 
integrals of this kind are known~\cite{dav}. 

\begin{figure}[t]
\centerline{
\psfig{figure=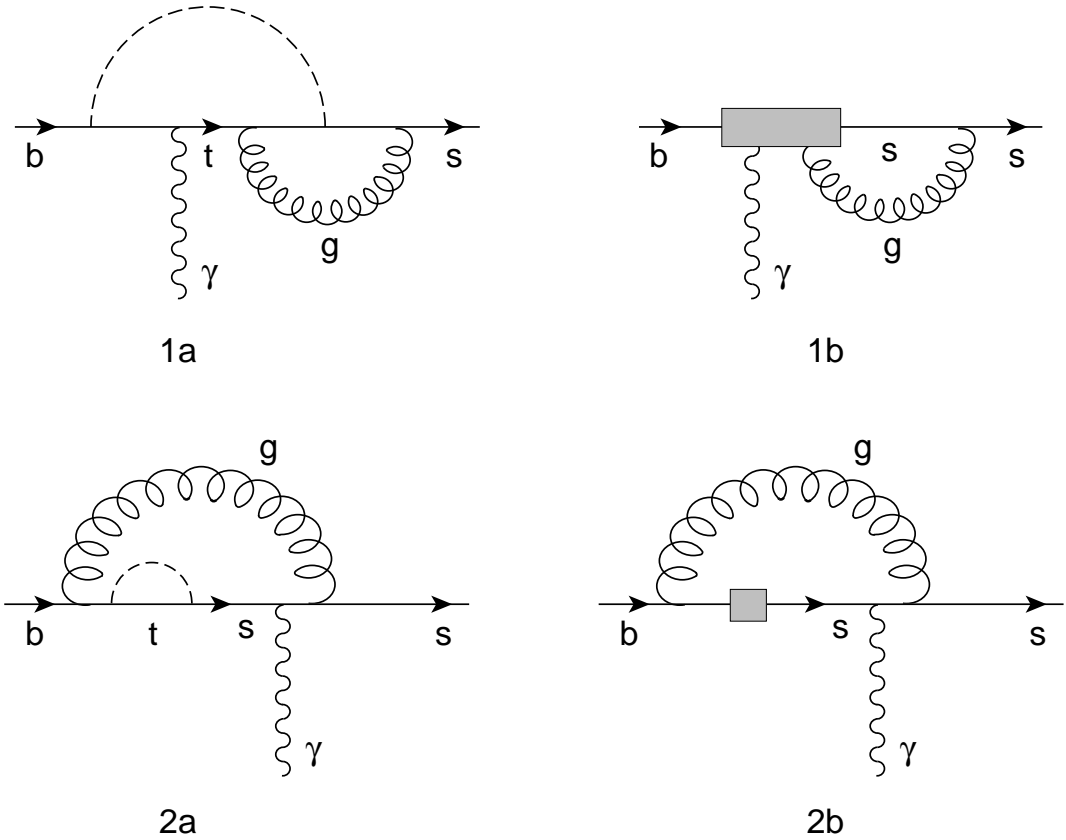,rheight=4.5in}
}
\caption{\sf Examples of the correspondence between diagrams in the
``full" theory and in the off-shell effective theory.} 
\label{fig1}
\end{figure}

The use of an off-shell configuration induces
in the result for the ``full" theory the appearance of 
terms that behave as $1/\ms^2$, $1/\mb^2$, $\log \, \ms^2$, $\log \, \mb^2$
as $\ms, \, \mb \rightarrow 0$. However, the
effective theory has the same infrared behaviour and there is a one-to-one
correspondence between  $\ms, \, \mb \rightarrow 0$
divergent contributions in
the diagrams of the ``full" and those of the effective theory.
It is worth reminding that this correspondence works in the off-shell
operator basis of eq.~(\ref{eq:basis}). 
In particular, diagrams like (1a) in 
fig.~\ref{fig1} have a counterpart in the effective theory, depicted in (1b),
that requires the introduction of a $ b \rightarrow s  g \gamma$ vertex.
Such a vertex is indeed present in the operators $\Gop{9},\, \Gop{10}, \,  
\Gop{12}, \, \Gop{13}$ while it does not appear in the on-shell operator
basis. The corresponding contribution can be worked out 
either through the use of the Ward identities that relate the 
1PI $ b \rightarrow s  \gamma$, $ b \rightarrow s g $ vertex to the
$ b \rightarrow s g \gamma$ one~\cite{SW}, or by direct computation of 
the Feynman rules from the relevant operators. In an analogous way, 
topology (2a)
requires a $b\to s$ transition vertex in the effective theory.
The leading behaviour of (2a) as $\ms, \mb  \rightarrow 0$ is 
${\cal O}(1/\ms^2)$, ${\cal O}(1/ \mb^2)$ and this
is  eliminated if the ``full" theory is renormalized on-shell~\cite{deshpande}
(see discussion in point 3 of section 2). 
Alternatively,  if the vanishing of the off-diagonal mass terms is not 
enforced, then the introduction of the dimension-four operators~\cite{simma}
\beq
\begin{array}{rl}
M_{1} = & \frac{i}{16\pi^2} \,\bar{s}_L  \not\!\! D \, b_L , 
\vspace{0.2cm} \\
M_{2} = & \frac{1}{16\pi^2} \mb \, \bar{s}_L \,  b_R , 
\end{array}
\eeq
in the diagram (2b) will provide such a cancellation. However, 
topology (2a) also shows an infrared 
logarithmic behaviour in the light quark masses. 
This cannot be eliminated through renormalization but it is actually cancelled 
when the operators $\Gop{9}$ and $\Gop{12}$ are inserted in diagram (2b).

It should be noticed that the cancellation of infrared 
$\ms,\, \mb$ contributions 
between ``full" and effective theory works before  
enforcing  the GIM mechanism. Therefore   the charm diagrams do not present
any additional difficulty. The charm contribution to $\Cp$ can be  
obtained by replacing $\mt$ with  $\mc$ and taking the limit
$\mc \rightarrow 0$.  After performing this operation one is left
with some $\log\, \mc$ terms. However, the effective theory contains
also the contribution of the operator $\Gop{2}$ that provides the
cancellation of these logarithmic terms through two-loop diagrams. 

As already said, the procedure for extracting  $\Cg$ is completely analogous,
the only difference being that  instead of introducing a
$ b \rightarrow s g \gamma$ vertex, one needs a $ b \rightarrow s g g$
vertex. This is present in the operators $\Gop{8},\, \Gop{9}, \,\Gop{12},
\, \Gop{13}, \, \Gop{14}$. Again, 
the infrared divergent contributions associated with the gluon cancel
diagram by diagram between the ``full" and the effective theory.

\section{Standard Model Results}
\label{sec:SM}
In this section we present the result for the SM. The analysis is performed
in the on-shell basis obtained by applying the EOM to the operators
$Q_9 \!-\! Q_{14}$. All the Wilson coefficients are now intended
in the on-shell basis, but for simplicity we drop the subscript $OS$.
The NLO expression for the
branching ratio of the inclusive radiative decay $B\to X_s \gamma$ 
has been presented in ref.~\cite{misiak1}:
\bea
\bsg \equiv &
BR(B\to X_s \gamma )&=BR(B \to X_c e \bar{\nu}_e)\left|
\frac{V_{ts}^* V_{tb}}{V_{cb}}\right|^2
\frac{6 \alpha_e}{\pi f(z)\kappa(z)}\frac{\bar m_b^2(\mu_b)}{m_b^2}
\nonumber \\
&& \times \left( |D|^2 + A \right)  \left( 1-\frac{\delta^{NP}_{SL}}{m_b^2}
+\frac{\delta^{NP}_{\gamma}}{m_b^2}+\frac{\delta^{NP}_{c}}{m_c^2}\right) ~.
\label{br}
\eea
Here and throughout the paper $m_q$ ($q=c,b,t$)
denotes the quark pole mass, while ${\bar m}_q$
denotes the $\overline{MS}$ running mass,
\beq
\bar m_q(m_q)=m_q\left[ 1-\frac{4}{3}\frac{\al (m_q)}{\pi}\right] ~.
\label{pole}
\eeq
Also $z=m_c^2/m_b^2$ and $f(z)$ and $\kappa (z)$ are the
phase-space factor and the QCD correction~\cite{cabib}
for the semileptonic decay,
\beq
\kappa(z)=1-\frac{2\al (\bar \mu_b)}{3\pi}\frac{h(z)}{f(z)}
\label{kap}
\eeq
\beq
f(z) = 1 - 8 z + 8 z^3 - z^4 - 12 z^2 \ln z 
\eeq
\bea
h(z) =
- (1-z^2) \left( \f{25}{4}- \f{239}{3} z + \f{25}{4} z^2 \right) 
+ z \ln z \left( 20 + 90 z - \f{4}{3} z^2 + \f{17}{3} z^3 \right) 
&& \nonumber \\
+ z^2 \ln^2 z \; ( 36 + z^2) 
+ (1-z^2) \left( \f{17}{3} - \f{64}{3} z + \f{17}{3} z^2 \right) \ln (1-z) 
&& \nonumber \\
- 4 ( 1 + 30 z^2 + z^4 ) \ln z \; \ln (1-z) 
- (1 + 16z^2 + z^4) \left[ 6 {\rm Li}_2(z) - \pi^2 \right] 
&& \nonumber \\
- 32 z^{3/2} (1+z) \left[ \pi^2 
     - 4 {\rm Li}_2(\sqrt{z}) +  4 {\rm Li}_2(-\sqrt{z}) 
     - 2 \ln z \; \ln \left( \f{1-\sqrt{z}}{1+\sqrt{z}} \right) \right]. 
&& 
\eea 
In eq.~(\ref{kap})
$\bar \mu_b$ is the renormalization scale relevant to the semileptonic
process, {\it a priori} different than the renormalization scale relevant
to the radiative decay, as stressed in ref.~\cite{buras2}.

In order to relate the quark decay rate to the actual hadronic process, we
have included the corrections obtained with the method of the 
Heavy-Quark Effective Theory (HQET), 
for a review, see {\it e.g.}~ref.~\cite{neubr}. 
In terms of
the contributions to a heavy hadron mass from kinetic energy ($\lambda_1$)
and chromo-magnetic interactions ($\lambda_2$), the $1/m_b^2$ corrections
to the semileptonic and radiative decays are
\beq
\delta_{SL}^{NP}=\frac{\lambda_1}{2}+\frac{3}{2}\lambda_2\left[
1-4\frac{(1-z)^4}{f(z)}\right]~,
\eeq
\beq
\delta_{\gamma}^{NP}=\frac{\lambda_1}{2}-\frac{9}{2}\lambda_2~.
\eeq
Notice that the dependence on the poorly-known parameter $\lambda_1$ drops
out in eq.~(\ref{br}). We take $\lambda_2=(m_{B^*}^2-m_B^2)/4
=0.12$ GeV$^2$, as extracted from
B-meson mass splitting. Voloshin~\cite{vol} has considered also 
$1/m_c^2$ non-perturbative corrections coming from interference between the
operators $O_2$ and $O_7$. With the correct sign found in ref.~\cite{isi},
these are given by 
\beq
\delta_{c}^{NP}=-\frac{1}{9}\frac{\lambda_2}{C^{(0)}_7}\left(C^{(0)}_2-
\frac{C^{(0)}_1}{6}\right)~,
\eeq
where $C^{(0)}_2$ and $C^{(0)}_7$ 
are the corresponding Wilson coefficients at the
LO computed at the scale $\mu_b$. 
It appears that higher-order terms in $1/m_c^2$ are 
suppressed~\cite{mcc}, and therefore non-perturbative corrections are
presumably under control.

In eq.~(\ref{br}), $A$ is the correction coming from the bremsstrahlung
process $b\to s\gamma g$, with a cutoff on the photon energy $E_\gamma
>(1-\delta)m_b/2$~\cite{ali,pott},
\clearpage
\bea
A &=& \left( e^{ -\al(\mu_b) \ln \delta (7 + 2 \ln \delta)/ 3 \pi} - 1 \right)
|C_7^{(0)eff}(\mu_b)|^2  \nonumber \\ 
&& + \frac{\al(\mu_b)}{\pi} 
\sum_{ \stackrel{i,j=1}{i \leq j}}^8
         C_i^{(0)eff}(\mu_b) C_j^{(0)eff}(\mu_b) f_{ij}(\delta).
\eea
The coefficients $f_{ij}(\delta)$, which have been computed in
ref.~\cite{ali,pott},
are defined as in ref.~\cite{misiak1}. 
To conform
to previous literature, we define a ``total" inclusive decay by choosing
$\delta=0.99$, although for comparison we will also present
results for $\delta=0.90$.
 The numerical values of the coefficients $f_{ij}(\delta)$,
for $\delta =0.99$ and $\sqrt{z}=0.29$, are\footnote{
We thank M. Misiak and N.~Pott
for pointing out to us  misprints in the expressions of $f_{12}$
in ref.~\cite{misiak1} and $f_{78}$ in 
ref.~\cite{pott}.} 
{\footnotesize
\beq
f_{ij} = \pmatrix{
 0.0009 & -0.0113 & -0.0035 & 0.0006 &-0.0459 & -0.0600 & -0.0030 &  0.0010\cr
        &  0.0340 &  0.0210 &-0.0035 & 0.2754 &  0.3599 &  0.0182 & -0.0061\cr
        &         &  0.0140 &-0.0047 & 0.3277 &  0.0666 &  0.0421 & -0.0140\cr 
        &         &         & 0.0088 &-0.0546 &  0.1570 & -0.0070 &  0.0023\cr 
        &         &         &        & 1.9369 &  0.9506 &  0.5926 & -0.1975\cr
        &         &         &        &        &  0.9305 &  0.0002 & -0.0001\cr 
        &         &         &        &        &         &  3.4211 &  0.3897\cr 
        &         &         &        &        &         &         &  3.2162  }
\eeq
}

Finally
\beq
D = C_7^{(0)eff}(\mu_b) + \frac{\al(\mu_b)}{4 \pi} \left\{ 
C_7^{(1)eff}(\mu_b) + \sum_{i=1}^8 C_i^{(0)eff}(\mu_b) \left[ 
r_i + \gamma_{i7}^{(0)eff} \ln \frac{m_b}{\mu_b} 
\right] \right\}~.
\label{dterm}
\eeq
The coefficients $r_i$, obtained from an NLO matrix element calculation,
have been calculated in ref.~\cite{wyler} for $i=1,2,7,8$. 
They can be found in ref.~\cite{misiak1}, together with the corresponding
values of $\gamma_{i7}^{(0)eff}$.
The dependence on $\mu_b$,
the renormalization
scale relevant for the $B\to X_s \gamma$ 
process,
 cancels out
in $D$ at the leading order in $\al$. The relations between the
Wilson coefficients at the low-energy scale $\mu_b$ and the high-energy 
scale $\muw$, obtained by renormalization group evolution, 
can be found in refs.~\cite{misiak1,buras2}. 
We are interested here in
the expressions of the Wilson coefficients at the scale $\muw$,
at which the ``full" theory is matched into an effective theory
with five quark flavours. In the SM model, they are given by
\beq
C^{eff}_i(\muw) = C^{(0)eff}_i(\muw) + \frac{\al(\muw)}{4 
\pi} C^{(1)eff}_i(\muw) 
\eeq
\beq
C_i^{(0)eff}(\muw) = C_i^{(0)}(\muw) = \left\{ \begin{array}{cc}
0 & \mbox{ for $i = 1,3,4,5,6$} \\
1 & \mbox{ for $i = 2$} \\
\vspace{0.2cm} 
F_i^{(1)}(x)
                                                  & \mbox{ for $i = 7,8$}, \\
\end{array} \right.
\label{wilslo}
\eeq
\beq
F_7^{(1)}(x)=\frac{x(7-5x-8x^2)}{24(x-1)^3}+\frac{x^2(3x-2)}{4(x-1)^4}\ln x
\label{f71}
\eeq
\beq
F_8^{(1)}(x)=\frac{x(2+5x-x^2)}{8(x-1)^3}-\frac{3x^2}{4(x-1)^4}\ln x
\label{f81}
\eeq
\beq
x=\frac{\bar m_t^2(\muw )}{\mw^2}.
\eeq
The NLO top-quark running mass at the scale $\muw$ is given by
\beq
\bar m_t (\muw )=\bar m_t (m_t) \left[ \frac{\al (\muw)}{\al (m_t )}
\right]^{\frac{\gamma_0^m}{2\beta_0}}
\left[ 1+ \frac{\al (m_t)}{4\pi}\frac{\gamma_0^m}{2\beta_0}\left(
\frac{\gamma_1^m}{\gamma_0^m}-\frac{\beta_1}{\beta_0}\right)
\left( \frac{\al (\muw )}{\al (m_t)}-1\right) \right]~,
\eeq
\beq
\beta_0=
11-\frac{2}{3}n_f~,~~~~
\beta_1=102-\frac{38}{3}n_f~,
\eeq
\beq
\gamma_0^m= 8~,~~~~\gamma_1^m=\frac{404}{3}-\frac{40}{9}n_f~,
\eeq
and it is related to the pole mass by eq.~(\ref{pole}). 
Here $n_f$ is the effective number of quark flavours.
We have computed
the NLO corrections using the method illustrated in sect.~3 and confirmed
the results of ref.~\cite{yao}\footnote{We are  grateful to
F.~Borzumati and C.~Greub for pointing out to us the omission of the 
logarithmic corrections to $C_{1,4}^{(1)eff}(\muw)$ in the preprint version 
of this work (see also ref.~\cite{borz}).}
\beq
C_i^{(1)eff}(\muw)  = \left\{ \begin{array}{cc}
15 + 6 \ln\frac{\muw^2}{\mw^2} & \mbox{ for $i = 1$} \\
0 & \mbox{ for $i = 2,3,5,6$} \\
E(x) - \frac{2}{3} +\frac{2}{3}\ln\frac{\muw^2}{\mw^2} & \mbox{ for $i = 4$}\\
G_i(x)+\Delta_i(x)\ln\frac{\muw^2}{\mw^2} & 
 \mbox{for $i = 
7,8$} , 
\end{array} \right.
\label{wilsnlo}
\eeq
\beq
E(x) = \frac{x (-18 +11
x + x^2)}{12 (x-1)^3} + \frac{x^2 (15 - 16 x + 4 x^2)}{6 (x-1)^4} \ln
x-\frac{2}{3} \ln x.
\eeq
\bea
G_7 (x) &=& \frac{-16 x^4 -122 x^3 + 80 x^2 -  8 x}{9 (x-1)^4} 
{\rm Li}_2 \left( 1 - \frac{1}{x} \right)
                  +\frac{6 x^4 + 46 x^3 - 28 x^2}{3 (x-1)^5} \ln^2 x 
\nonumber \\ &&
                  +\frac{-102 x^5 - 588 x^4 - 2262 x^3 + 3244 x^2 - 1364 x +
208} {81 (x-1)^5} \ln x
\nonumber \\ &&
                  +\frac{1646 x^4 + 12205 x^3 - 10740 x^2 + 2509 x - 436}
{486 (x-1)^4} 
\label{c71eff}
\vspace{0.2cm} \\
G_8(x) &=& \frac{-4 x^4 +40 x^3 + 41 x^2 + x}{6 (x-1)^4} 
{\rm Li}_2 \left( 1 - \frac{1}{x} \right)
                  +\frac{ -17 x^3 - 31 x^2}{2 (x-1)^5} \ln^2 x 
\nonumber \\ &&
                  +\frac{ -210 x^5 + 1086 x^4 +4893 x^3 + 2857 x^2 - 1994 x
+280} {216 (x-1)^5} \ln x
\nonumber \\ &&
        +\frac{737 x^4 -14102 x^3 - 28209 x^2 + 610 x - 508}{1296 (x-1)^4}
\label{c81eff}
\eea
\beq
\Delta_7(x)=\frac{208-1111x+1086x^2+383x^3+82x^4}{81(x-1)^4}
+\frac{2x^2(14-23x-3x^2)}{3(x-1)^5}\ln x
\label{del7}
\eeq
\beq
\Delta_8(x)=\frac{140-902x-1509x^2-398x^3+77x^4}{108(x-1)^4}
+\frac{x^2(31+17x)}{2(x-1)^5}\ln x
\label{del8}
\eeq
The $\muw$ dependence in all terms of order
$\al$ in eq.~(\ref{dterm}) cancels out because eqs.~(\ref{del7}) and
(\ref{del8}) satisfy the relation
\beq
\Delta_i=\gamma_0^mx\frac{\partial C_i^{(0)}}{\partial x}+\frac{1}{2}
\sum_{j=1}^8\gamma_{ji}^{(0)eff}C_j^{(0)}
\label{canc}
\eeq


{\footnotesize
\begin{table}
\centering
\begin{tabular}{|c|c|c|c|c|c|c|c|c|c|}
\hline
$\mu_b$ & $\bar \mu_b$ & $\muw$ & $m_t$ & $ _{\left| \frac{V_{tb} 
V^\star_{ts}}{V_{cb}}
\right|^2} $ & $m_c/m_b$ & $m_b-m_c$ & $BR_{SL}$ &
$\alpha_s(M_Z)$ & $\alpha_e^{-1}$ \\
 \mbox{[GeV]} & [GeV] & [GeV] & [GeV] & & & [GeV] & & & \\
\hline
\multicolumn{10}{|c|}{Central values}\\
4.8 & 4.8 & 80.33 & 175 & 0.95 & 0.29 & 3.39 & 0.1049 & 0.118 & 130.3\\
\hline
\multicolumn{10}{|c|}{Errors/Ranges}\\
$^{+4.8}_{-2.4}$ & $^{+4.8}_{-2.4}$ & $^{+80.33}_{-40.16}$ & $\pm 5$ &
$\pm 0.03$ & $\pm 0.02$ & $\pm0.04$ & $\pm 0.0046$ & $\pm 0.003$ & $\pm 2.3$\\
\hline
\multicolumn{10}{|c|}{Uncertainty in $\bsg$ (LO)}\\
$_{+24}^{-19}\%$ & - & $^{+9.3}_{-9.7}\%$ & $\pm 1.6\%$ & $\pm 3.2\%$ &
$^{+8.5}_{-7.3}\%$ & $^{-0.07}_{+0.08}\%$ & $\pm 4.4\%$ 
& $^{+2.3}_{-2.2}\%$ & $^{-1.7}_{+1.8}\%$ \\
\hline
\multicolumn{10}{|c|}{Uncertainty in $\bsg$ (NLO)}\\
$^{+0.0}_{-3.5}\%$ & $^{-1.4}_{+2.1}$  
& $^{-1.1}_{+2.1}\%$ & $\pm 1.4\%$ & $\pm 3.2\%$ &
$^{+4.5}_{-3.8}\%$ & $\mp 0.4\%$ &
$\pm 4.4\%$ & $^{+2.9}_{- 2.8}\%$ & $^{-1.7}_{+ 1.8}\%$ \\
\hline
\end{tabular}
\caption{{\sf The input parameters entering the calculation of $\bsg$,
their central values, their errors, their influence on the uncertainty
in $\bsg$ at the LO and NLO. The other constants entering the calculation
are $\mw=80.33$ GeV, $\lambda_2=0.12$ GeV$^2$, $m_s/m_b=50$, and $\delta=0.99$.
}}
\end{table}
}

The values of the parameters chosen in our numerical analysis are shown 
in table 1. The value of $\al (M_Z)$, obtained from an average of low-
and high-energy measurements, is taken from ref.~\cite{pdg}. From this
we calculate $\al$ at a generic scale $\mu$ with the NLO result
\beq
\al (\mu)=\frac{\al (M_Z)}{v}\left( 1-\frac{\beta_1}{\beta_0}
\frac{\al (M_Z)}{4\pi}\frac{\ln v}{v}\right) ~~~~~
v=1+\beta_0\frac{\al (M_Z)}{2\pi}\ln \frac{\mu}{M_Z}~.
\eeq

The top quark mass is obtained by combining the results by
CDF~\cite{cdf}  ($m_t=176.8\pm 4.4^{stat}\pm 4.8^{syst}$ GeV) and D0~\cite{d0}
($m_t=173.3\pm 5.6^{stat}\pm 6.2^{syst}$ GeV). The difference between
the bottom and charm mass is obtained from the HQET relation between the
quark and hadron masses, $m_b-m_c=3.39\pm 0.03\pm 0.03$ GeV~\cite{neub2}, where
the first error is from the uncertainty in $\lambda_1$ and the second one
from higher-order corrections. The ratio $m_c/m_b$ is chosen in the
range $0.29\pm 0.02$, which corresponds to $m_b=4.77\pm0.15$ GeV. 

The uncertainty on the electromagnetic coupling $\alpha_e$ corresponds
to the arbitrariness of choosing the renormalization scale between $m_b$
and $\mw$, as QED corrections have not been included. 

The experimental situation on the semileptonic branching ratio $\bsl$ is
rather controversial (for a review, see ref.~\cite{richman}). The 
two measurements at the $\Upsilon (4S)$ give $\bsl =(10.49\pm 0.17
\pm 0.43)$\% (CLEO~\cite{cleosl}) and $\bsl =(9.7\pm 0.5
\pm 0.4)$\% (ARGUS~\cite{argussl}).
On the other hand, the average of the four LEP experiments 
gives~\cite{lepwg,feig}
$\bsl^Z=(10.95\pm 0.13\pm 0.29)$\%. However the $B$ samples at the $Z^0$ also
contain $B$ hadrons which have typically shorter lifetimes than $B$ mesons.
Following Neubert~\cite{neub2}, we can correct $\bsl^Z$ using the average 
lifetime
of the corresponding $B$ mixture and obtain
$
\bsl =(11.23\pm 0.34)\%$ at $Z^0$.
The low- and high-energy  results 
show a certain discrepancy. However, a reanalysis of the LEP results is
finding new systematic errors and a new dedicated analysis is in 
progress~\cite{feig}. For this reason we prefer to use only the CLEO result.

In terms of the Wolfenstein parameters $\lambda$, $\rho$, and $\eta$ related
to the standard parametrisation~\cite{pdg} by
\beq
s_{12}=\lambda ,~~~\frac{s_{13} e^{-i\delta}}{s_{23}}=\lambda (\rho -i\eta )
\eeq
and using unitarity, the relevant CKM factor can be written as
\beq
\left|
\frac{V_{ts}^* V_{tb}}{V_{cb}}\right|^2=1+(2\rho -1)\lambda^2~.
\eeq
For $\lambda =0.221$ and $\rho$ in the range between $-0.3$ and 0.3, we obtain
the value shown in table 1. 
The new physics contributions that we will consider in sect.~5 can modify
the values of $\rho$ obtained from fits of different flavour-physics
measurements. However, since we allow a rather large range for $\rho$, these
effects are included in the error.

To estimate higher-order effects,
we vary 
the different renormalization scales $\mu_b$,
$\bar \mu_b$, and $\muw$
in the ranges given in table 1,
which correspond to  shifts in the logarithms of the scale by an amount
$\pm \ln 2$ around their central values $\mu_b =\bar \mu_b=m_b$,
$\muw =\mw$. 
In the NLO calculation we have dropped all explicit
${\cal O}(\al^2)$ terms in eq.~(\ref{br}). To allow a better
comparison, we also
perform the calculation in LO approximation, where ${\cal O}(\al )$ terms are
set to zero and $\al$ is evolved using the one-loop renormalization
group equation.

The result of the numerical evaluation of eq.~(\ref{br}) is
\beq
\bsg = (2.66\pm 0.74)\times 10^{-4} ~~~({\rm LO})
\label{lo}
\eeq
\bea
\bsg &=& (3.62\pm 0.33)\times 10^{-4}  ~~~({\rm NLO},\delta=0.99)\nonumber\\
&&(3.54\pm 0.32)\times 10^{-4}  ~~~({\rm NLO},\delta=0.90)
\label{nlo}
\eea
The error is obtained by adding in quadratures the errors due to the
different input parameters shown in table 1. If the error is asymmetric, we
always choose the larger uncertainty.
The $\mu_b$ scale dependence dominates the error at the LO,
but it is
drastically reduced at the NLO. This is illustrated in fig.~2. The result
in eq.~(\ref{nlo}) is in agreement with those presented in 
refs.~\cite{misiak1,buras2}.
The slightly different $\mu_b$
dependence we are finding comes from our choice of defining $\bar m_b$ in
eq.~(\ref{br}) at the scale $\mu_b$ (where the operator $Q_7$ is defined),
instead of $m_b$ as in refs.~\cite{misiak1,buras2}. Of course this just
corresponds
to a higher-order effect, and this ambiguity is taken into account by the
theoretical uncertainty.

Given the controversial
situation in $\bsl$, it
is useful to quote also the prediction for the ratio $\bsg /\bsl$
(for $\delta=0.99$)
\beq
\bsg /\bsl = (2.54\pm 0.70)\times 10^{-3} ~~~({\rm LO})
\eeq
\beq
\bsg /\bsl = (3.45\pm 0.27)\times 10^{-3}  ~~~({\rm NLO})
\label{nlorat}
\eeq

\begin{figure}[t]
\vspace{-7cm}
\centerline{
\psfig{figure=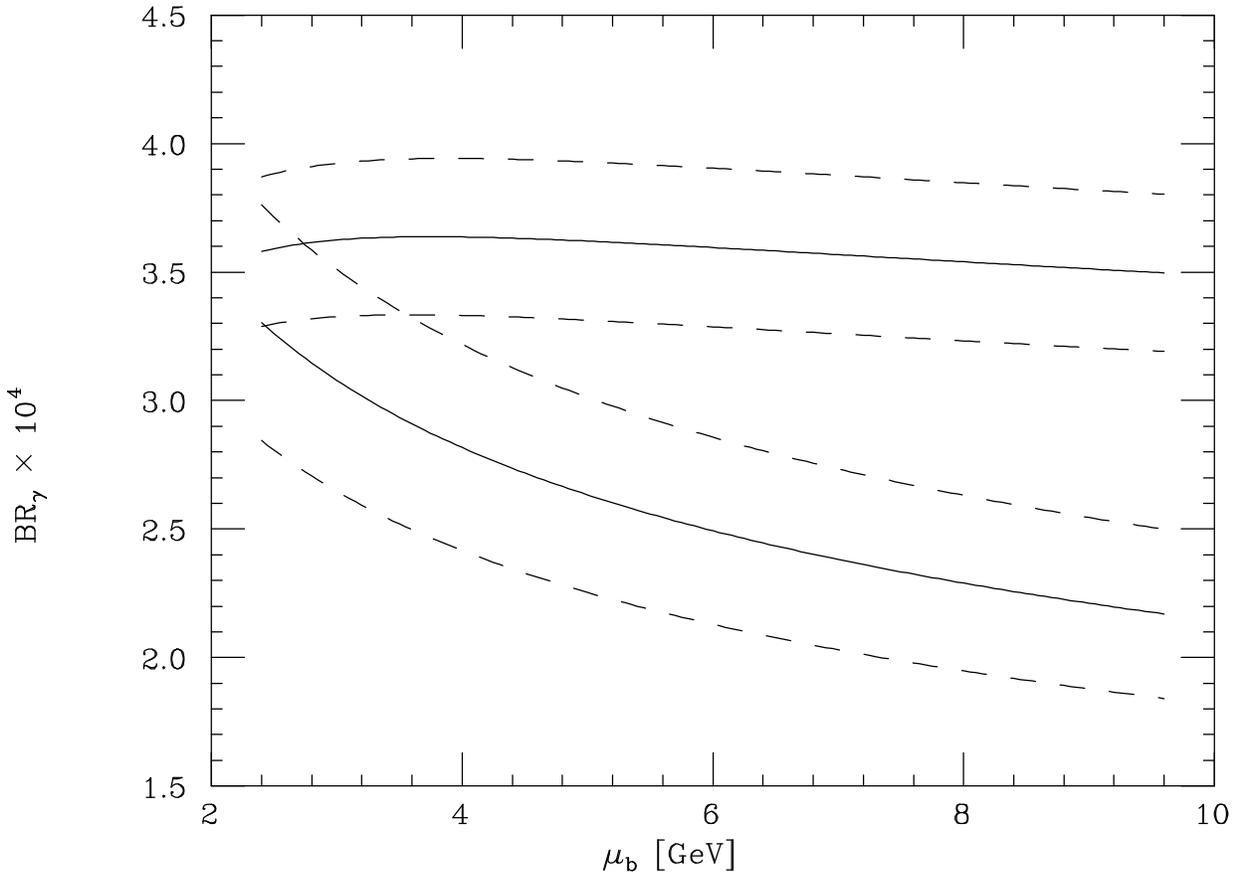,rheight=7.1 in}
  }
\caption{\sf The $\mu_b$ scale dependence of $\bsg(\delta=0.99)$
at LO (lower curves) 
and NLO (upper curves). The solid line represents the case in which all
input parameters are fixed to their central values.
The dashed lines represent the errors coming
from all
input parameters, other than $\mu_b$, added in quadratures.
}
\label{fig2}
\end{figure}

\section{Two-Higgs Doublet Models}
\label{sec:2HDM}
In this section we investigate the prediction for $\bsg$ in models with
two Higgs doublets. The new ingredient with respect to the SM is the
presence of a charged Higgs scalar particle which gives new contributions
to the Wilson coefficients of the effective theory. 
 The interaction
between quarks and the charged Higgs $H^\pm$ is defined by the Lagrangian
\beq
{\cal L}= (2\sqrt{2}G_F)^{1/2}\sum_{i,j=1}^3\bar u_i
\left( A_u m_{u_i}V_{ij}\frac{1-\gamma_5}{2}-
A_d V_{ij}m_{d_i}\frac{1+\gamma_5}{2}\right) d_j H^++{\rm h.c.}
\eeq
Here $i,j$ are generation indices, $m_{u,d}$ are quark masses, and $V$ is the
CKM matrix. We are considering only models
in which FCNCs vanish at tree level~\cite{nat}. 
Therefore we have two possibilities
which are parametrised by the model-dependent coefficients $A_u$ and $A_d$.
The first possibility, referred to as Model I, is that both up and down quarks
get their masses from the same Higgs doublet $H_2$, and
\beq
A_u=A_d=1/\tan \beta ~,
\eeq
where $\tan \beta=\langle H_2^0\rangle / \langle H_1^0\rangle$. The case of
Model II is that up quarks get their masses from Yukawa couplings to $H_2$,
while down quarks get masses from couplings to $H_1$, and
\beq
A_u=-1/A_d=1/\tan \beta ~.
\eeq
The latter 
case is particularly interesting because it is realized in the minimal
supersymmetric extension of the SM.

In the language of the effective theory, it is easy to include the 
charged-Higgs effect. We just need to compute the new
Wilson coefficients at the matching scale $\muw$, since they contain all
the relevant 
ultraviolet information. At the LO, the charged-Higgs contribution, to
be added to eq.~(\ref{wilslo}) is given by~\cite{chwil}
\beq
\delta C_{i}^{(0)eff}(\muw) =0 ~~~~~i=1,...,6
\eeq
\beq
\delta C_{7,8}^{(0)eff}(\muw) =\frac{A_u^2}{3}
F_{7,8}^{(1)}(y)-A_uA_dF_{7,8}^{(2)}(y)
\eeq
\beq
F_7^{(2)}(y)=\frac{y(3-5y)}{12(y-1)^2}+\frac{y(3y-2)}{6(y-1)^3}\ln y
\eeq
\beq
F_8^{(2)}(y)=\frac{y(3-y)}{4(y-1)^2}-\frac{y}{2(y-1)^3}\ln y
\eeq
\beq
y=\frac{\bar m_t^2(\muw )}{M_H^2}~,
\eeq
and $F_{7,8}^{(1)}$ are defined in eqs.~(\ref{f71})--(\ref{f81}). 

At
the NLO, the charged-Higgs contributions to the Wilson coefficients are
\beq
\delta C_{i}^{(1)eff}(\muw) =0 ~~~~~i=1,2,3,5,6
\eeq
\beq
\delta C_{4}^{(1)eff}(\muw) =E^H(y) 
\eeq
\bea
\delta C_7^{(1)eff}(\muw) &=& G_7^H(y) + 
       \Delta_7^H(y) \ln\frac{\muw^2}{M_H^2}-\frac49 E^H(y) \nonumber\\
\delta C_8^{(1)eff}(\muw) &=& G_8^H(y) + 
       \Delta_8^H(y) \ln\frac{\muw^2}{M_H^2} -\frac16 E^H(y) 
\eea
with
\bea
G_7^H(y) &= &  
A_dA_u\frac{4}{3}y\left[ \frac{4(-3+7y-2y^2)}{3(y-1)^3}{\rm Li}_2 
\left( 1 - \f{1}{y} \right)+
\frac{8-14y-3y^2}{3(y-1)^4}\ln^2y \right.
\nonumber \\ &&
\left. +\frac{2(-3-y+12y^2-2y^3)}{3(y-1)^4}\ln y
+\frac{7-13y+2y^2}{(y-1)^3}\right]
\nonumber \\ &&
+A^2_u\frac{2}{9}y\left[ \frac{y(18-37y+8y^2)}{(y-1)^4}{\rm Li}_2 
\left( 1 - \f{1}{y} \right)+
\frac{y(-14+23y+3y^2)}{(y-1)^5}\ln^2y \right.
\nonumber \\ &&
 +\frac{-50+251y-174y^2-192y^3+21y^4}{9(y-1)^5}\ln y
\nonumber \\ &&
\left . +\frac{797-5436y+7569y^2-1202y^3}{108(y-1)^4}\right]
\label{higceq}
\eea
\bea
\Delta_7^H(y) &= & A_dA_u\frac{2}{9}y\left[ \frac{21-47y+8y^2}{
(y-1)^3}+\frac{2(-8+14y+3y^2)}{(y-1)^4}\ln y \right]
\nonumber \\ &&
+A_u^2\frac{2}{9}y\left[ \frac{-31-18y+135y^2-14y^3}{
6(y-1)^4}+\frac{y(14-23y-3y^2)}{(y-1)^5}\ln y \right]
\eea
\bea
G_8^H(y) &= & 
A_dA_u\frac{1}{3}y\left[ \frac{-36+25y-17y^2)}{2(y-1)^3}{\rm Li}_2 
\left( 1 - \f{1}{y} \right)+
\frac{19+17y}{(y-1)^4}\ln^2y \right.
\nonumber \\ &&
\left. +\frac{-3-187y+12y^2-14y^3)}{4(y-1)^4}\ln y
+\frac{3(143-44y+29y^2)}{8(y-1)^3}\right]
\nonumber \\ &&
+A^2_u\frac{1}{6}y\left[ \frac{y(30-17y+13y^2)}{(y-1)^4}{\rm Li}_2 
\left( 1 - \f{1}{y} \right)-
\frac{y(31+17y)}{(y-1)^5}\ln^2y \right.
\nonumber \\ &&
 +\frac{-226+817y+1353y^2+318y^3+42y^4}{36(y-1)^5}\ln y
\nonumber \\ &&
\left. +\frac{1130-18153y+7650y^2-4451y^3}{216(y-1)^4}\right]
\eea
\bea
\Delta_8^H(y) &= &  A_dA_u\frac{1}{3}y\left[ \frac{81-16y+7y^2}{
2(y-1)^3}-\frac{19+17y}{(y-1)^4}\ln y \right]
\nonumber \\ &&
+A_u^2\frac{1}{6}y\left[ \frac{-38-261y+18y^2-7y^3}{
6(y-1)^4}+\frac{y(31+17y)}{(y-1)^5}\ln y \right]
\eea
\beq
E^H(y)=A_u^2\left[ \frac{y(16-29y+7y^2)}{36(y-1)^3}+\frac{y(3y-2)}{6(y-1)^4}
\ln y \right] ~.
\eeq
The $\mu_\smallw$ dependence cancels in terms of 
${\cal O}(\al )$, since the $H^\pm$ contribution satisfies eq.~(\ref{canc}).

\begin{figure}[t]
\vspace{-7cm}
\centerline{
\psfig{figure=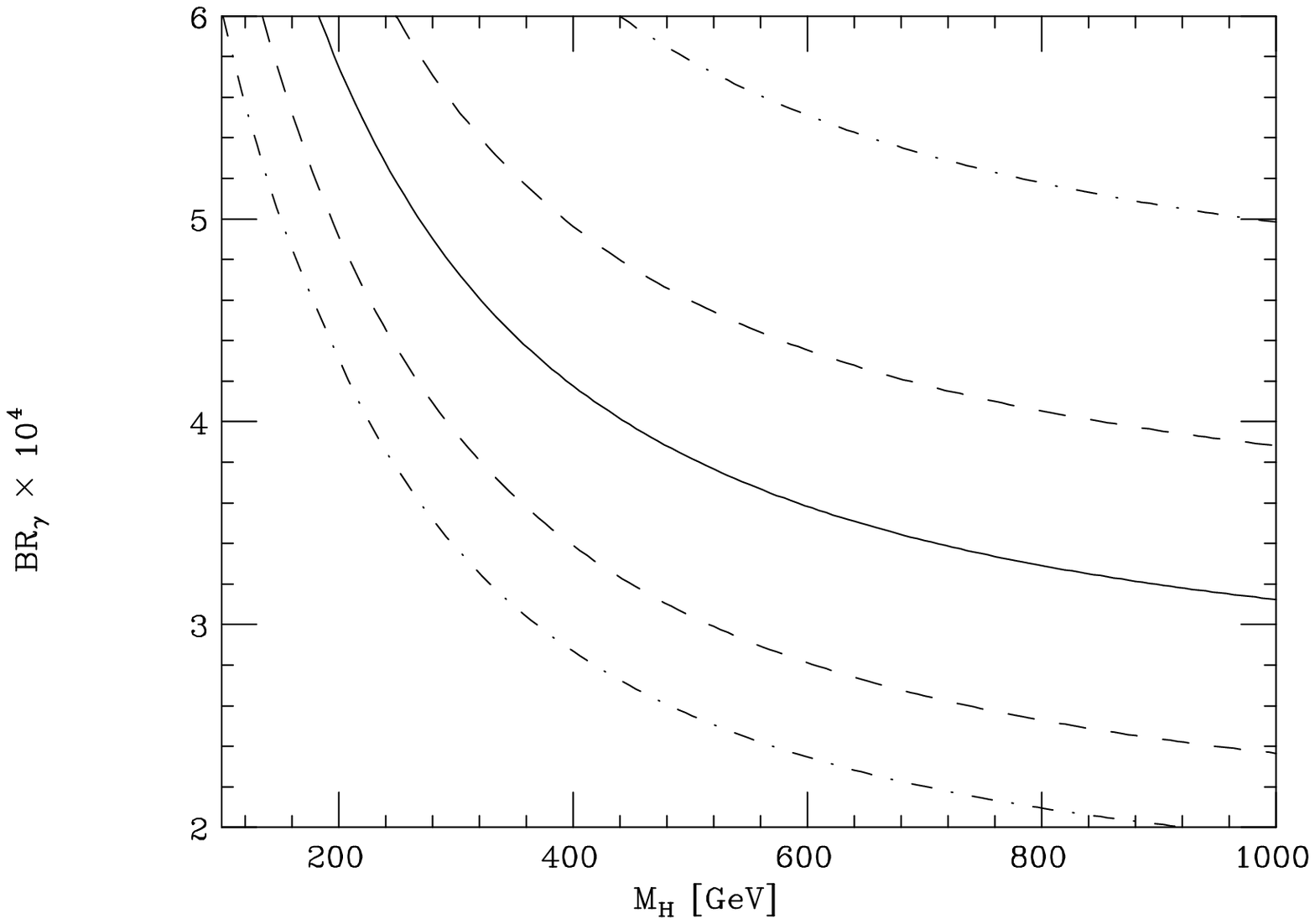,rheight=7.1in}
  }
\caption{\sf Predicted $\bsg(\delta=0.99)$ in Model II as a function of the
charged-Higgs mass $M_H$, for $\tan \beta =2$ at LO (solid line). 
The dashed lines
represent the uncertainty obtained by adding in quadratures the errors
coming from the different input parameters. The dot-dashed lines define
the uncertainty range with the ``scanning" method (see text).
} 
\label{fig3}
\end{figure}

\begin{figure}[t]
\vspace{-7cm}
\centerline{
\psfig{figure=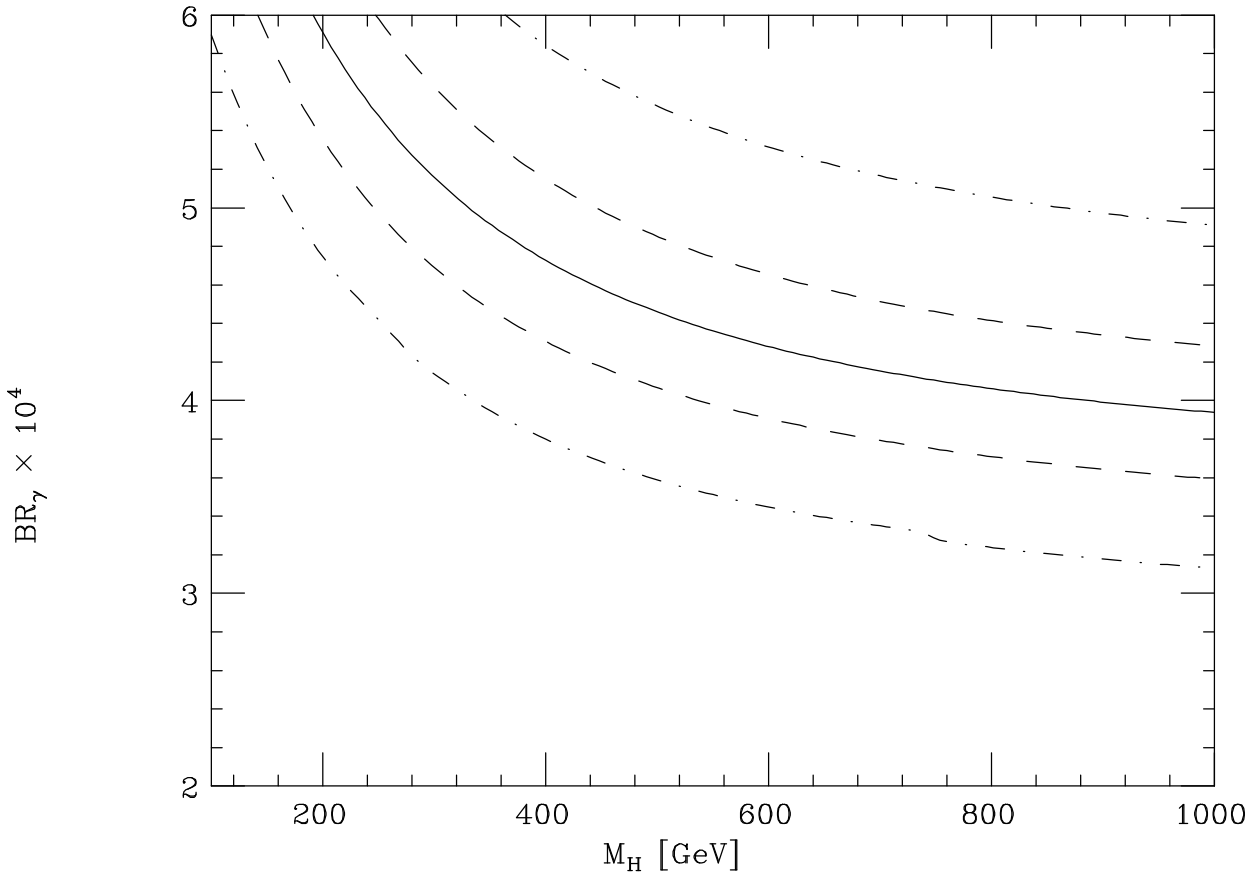,rheight=7.1in}
  }
\caption{\sf Predicted $\bsg(\delta=0.99)$ in Model II as a function of the
charged-Higgs mass $M_H$, for $\tan \beta =2$ at NLO (solid line). 
The dashed lines
represent the uncertainty obtained by adding in quadratures the errors
coming from the different input parameters. The dot-dashed lines define
the uncertainty range with the ``scanning" method (see text).}
\label{fig4}
\end{figure}

In figs.~\ref{fig3} and \ref{fig4}
 we show the predicted value of $\bsg$ in Model II
as a function of the
charged-Higgs mass $M_H$, for $\tan \beta =2$ at LO and NLO with 
$\delta=0.99$. 
As apparent from these
figures, the decoupling at large
$M_H$ occurs very slowly. Moreover, the $H^\pm$ contribution always
increases the SM value of $\bsg$. Since the SM prediction is
already quite higher than the CLEO measurement, one can obtain very
stringent bounds on $M_H$~\cite{higlo}. These bounds are very sensitive on
the errors of the theoretical prediction. Improving the calculation to
the NLO has important effects on these bounds, since the theoretical error
is significantly reduced, as shown in figs.~3 and 4.

Because of the large sensitivity on errors, we
will use two different methods to define the bounds on $M_H$.
We define a ``Gaussian" standard deviation by adding different input errors
in quadratures.
We also 
define a ``scanning" method of estimating errors by varying independently
all input parameters within the ranges defined in table 1. Figures~3 and
4 show the errors on the theoretical prediction 
using one ``Gaussian" standard deviation (dashed lines), or
using the ``scanning" method
(dot-dashed lines).
If we apply the
``scanning" method to the SM, we obtain (for $\delta=0.99$)
\beq
\bsg = (2.66 ^{+1.78}_{-1.09})\times 10^{-4} ~~~({\rm LO})
\eeq
\beq
\bsg = (3.62 ^{+0.92}_{-0.77})\times 10^{-4}  ~~~({\rm NLO})
\eeq
\beq
\bsg /\bsl = (2.54 ^{+1.52}_{-0.97})\times 10^{-3} ~~~({\rm LO})
\eeq
\beq
\bsg /\bsl = (3.45 ^{+0.69}_{-0.61})\times 10^{-3}  ~~~({\rm NLO})
\eeq
This has to be compared with the ``Gaussian" standard deviation error given
in eqs.~(\ref{lo})--(\ref{nlorat}).

\begin{figure}[t]
\vspace{-7cm}
\centerline{
\psfig{figure=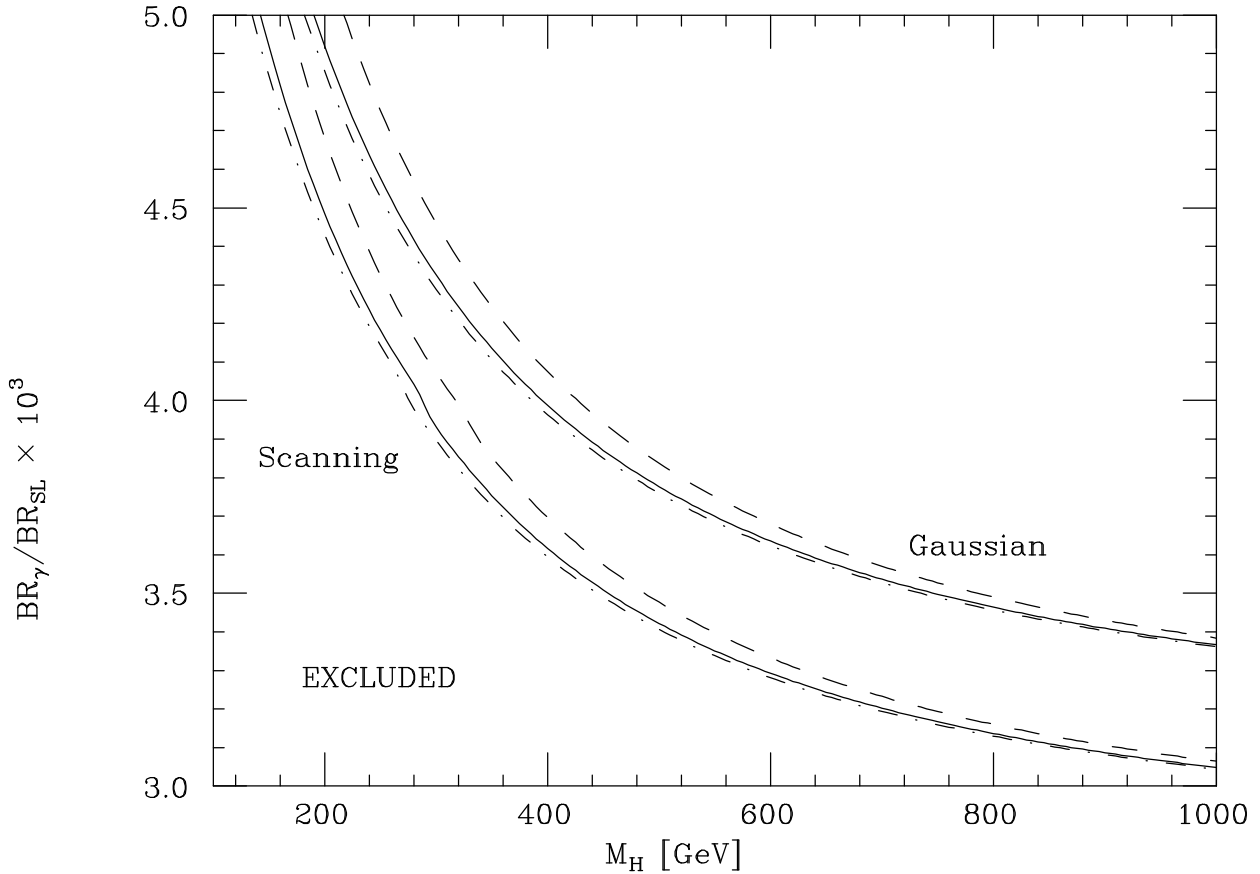,rheight=7.1in}
  }
\caption{\sf Exclusion regions in the plane $\bsg /\bsl$ versus $M_H$
for $\tan \beta =$ 1 (dashed lines), 2 (solid lines), 5 (dot-dashed lines)
obtained by our NLO calculation with $\delta=0.90$. 
The upper lines correspond to
a theoretical uncertainty defined by one ``Gaussian" standard deviation,
and the lower lines to
the ``scanning" method (see text). 
}\label{fig5}
\end{figure}

Next, we exclude a value of $M_H$ if the minimum of $\bsg$ inside the
``Gaussian" (or ``scanning") 
band lies above the present CLEO 95\% CL upper bound $\bsg < 4.2
\times 10^{-4}$~\cite{cleo}. 
The corresponding limits on $M_H$ are shown in table~2, 
for different values
of $\tan \beta$. If we restrict our consideration to the case
$\tan \beta >1$,
the dependence on $\tan \beta$ is very mild,
and the limit practically saturates for $\tan \beta \gappeq 2$. Notice
that, for a ``Gaussian" error,
going from the LO to
the NLO calculation allows an improvement of the limit on $M_H$ from
about 260
GeV to about 370 GeV, for large $\tan \beta$, even in the more conservative 
case of $\delta=0.90$. 
Had we neglected the NLO Higgs matching condition, but kept the remaining
NLO corrections (as done in ref.~\cite{pok}), we would find a much
stronger limit $M_H>460$ GeV ($\tan \beta =5$, ``Gaussian" error).

\begin{table}
\centering
\begin{tabular}{|c|c|c|c|c|}
\hline
$\tan \beta$ & theor. error & $M_H$ (LO) & $M_H$ (NLO)&$M_H$ (NLO)\\
 & & [GeV] & [GeV]($\delta=0.99$) & [GeV] ($\delta=0.90$)\\
\hline
1 & ``Gaussian" & 300 & 473 & 408\\
2 & ``Gaussian" & 268 & 440 & 377\\
5 & ``Gaussian" & 258 & 430 & 368\\
\hline
1 & ``scanning" & 240 & 316 & 276 \\
2 & ``scanning" & 208 & 287 & 249 \\
5 & ``scanning" & 198 & 278 & 242 \\
\hline
\end{tabular}
\caption{{\sf Lower bounds on the charged-Higgs mass $M_H$ from 
the CLEO 95\% CL limit $\bsg < 4.2 \times 10^{-4}$, using
LO or NLO
calculations. The theoretical uncertainty is estimated either with 
one ``Gaussian" standard deviation, with input parameter errors
added in quadratures, or with the
 ``scanning" method, by varying independently the input parameters within
their range. The NLO results are presented for two different values of the 
cut on the photon energy $\delta$.
}}
\end{table}


It may seem disturbing to integrate out the top quark, the $W$, and a
charged Higgs as heavy as 300 GeV or more at the same scale $\muw$. 
However, since
we have completed the NLO calculation, the residual dependence on $\muw$ is
very small.
A resummation of the leading logarithms of $M_H/m_W$ has been performed in 
ref.~\cite{anl}.

\begin{figure}[t]
\vspace{-7cm}
\centerline{
\psfig{figure=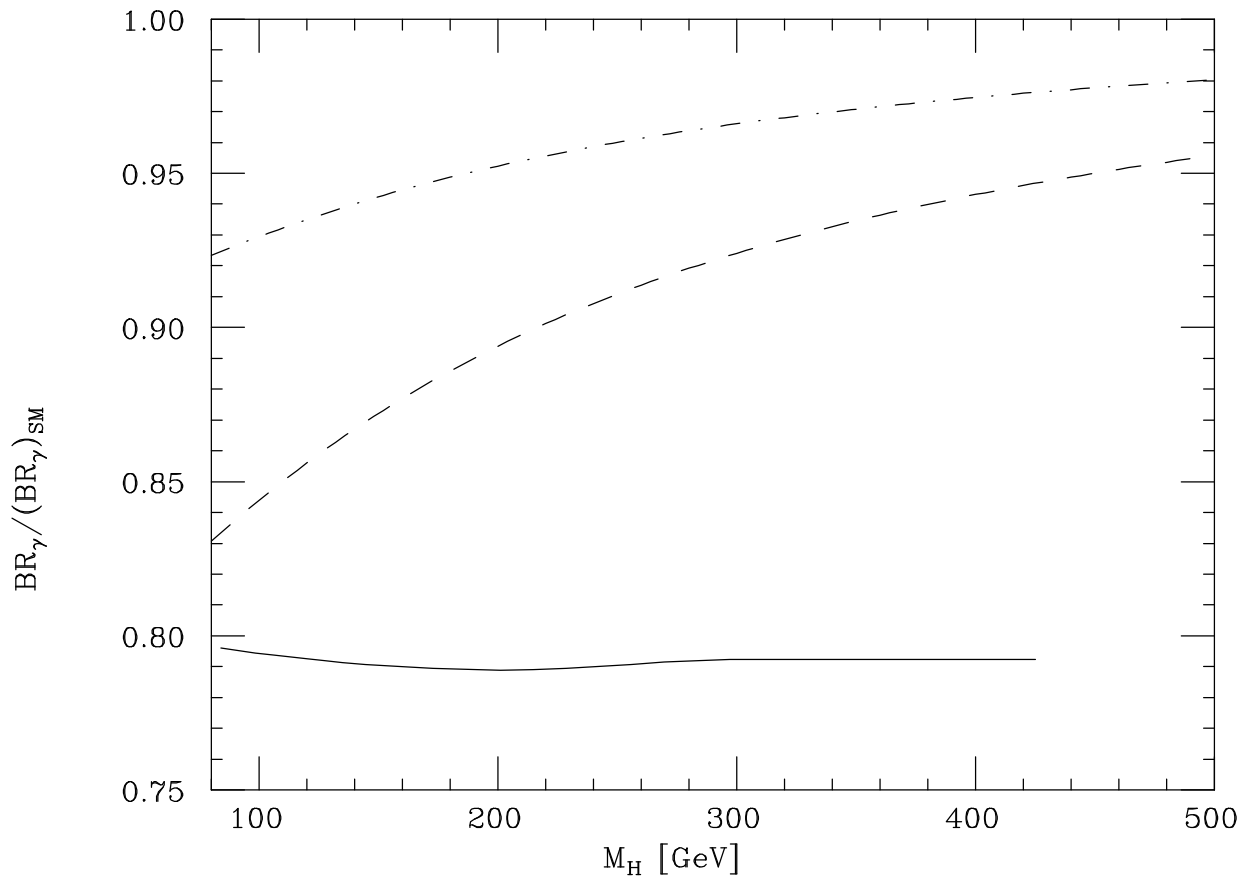,rheight=7.12in}
  }
\caption{\sf The ratio between $\bsg(\delta=0.90)$ predicted in Model I and
the SM value,
for the minimum $\tan \beta$ allowed by the $R_b$ measurement (solid line),
for $\tan \beta =2$ (dashed line) and  $\tan \beta =3$ (dot-dashed line).
The solid line is interrupted at $\tan \beta =1$.
}
\label{fig6}
\end{figure}

Certainly one of the reasons we are able to obtain such strong bounds on
$M_H$ is the poor agreement between SM prediction and CLEO measurement.
In view of the preliminary ALEPH result on $\bsg$ and the controversial
situation on $\bsl$, it is more
appropriate to present our results in a fashion which is independent of
the present measurements. This is done in fig.~5,
which shows the bound
on $M_H$ according to the ``Gaussian" and 
``scanning" methods, for any given value of the
experimental upper bound on $\bsg/\bsl$ in the more conservative case
$\delta=0.90$. 
The region below the curves is excluded by our theoretical calculation.
As future measurements on $\bsg/\bsl$ improve,
on can simply read off the limit on $M_H$ from the lines of fig.~5.

It should be remarked that, although Model II corresponds to the 
supersymmetric Higgs sector, these bounds do not directly apply to
the supersymmetric extension of the SM. The reason is that 
contributions coming from other new particles can partly compensate the
effect of the charged Higgs. In particular, in the supersymmetric limit, there
is an exact cancellation of different contributions~\cite{io}. In the
realistic case of broken supersymmetry, this cancellation is spoiled but,
if charginos and stops are light, it may still be partially effective.
Nevertheless, the bound presented here severely restricts the allowed
region of parameter space.

In the case of Model I, the charged-Higgs contribution reduces the 
value of $\bsg$ and therefore we cannot obtain any significant bound. On
the other hand, it is interesting that new physics effects can
bring the prediction for $\bsg$ closer to the CLEO value. 

A significant effect can only be expected for small $\tan \beta$, since
in Model I all charged-Higgs contributions vanish in the limit of
large $\tan \beta$, as $\tan^{-2} \beta$. When $\tan \beta$ is reduced, the
top Yukawa coupling grows, and one eventually is limited by new contributions
to $R_b$, $B^0$-$\bar B^0$ mixing, and occurrence of a Landau pole at
intermediate energies (see {\it e.g.} ref.~\cite{lim}). 
We have checked that the strongest experimental
lower bound on $\tan \beta$ comes from $R_b$, the ratio between the
bottom and hadronic $Z^0$ widths\footnote{We thank C.~Wagner for providing
us with a numerical calculation of $R_b$.}. Using the 95\% CL lower limit
$R_b>0.2155$ obtained from a combined LEP+SLD analysis ($R_b=0.2170
\pm0.009$~\cite{lepwg}) and the smallest top-quark mass value inside our range
($m_t=170$ GeV), we derive a lower bound on $\tan \beta$ of 1.8, 1.4, 1.0
for $M_H=85$, 200, and 425 GeV, respectively. For these  values of
$\tan \beta$ and $M_H$, the $B^0$-$\bar B^0$ mixing parameter is increased
by about 30\% with respect to the SM. Such an increase is compatible
with experiments, given that the theoretical uncertainty on the hadronic
parameter $F_B^2B_B$ is at least 40{\%}. In correspondence to this minimum
value of $\tan \beta$ we obtain the maximum reduction factor of $\bsg$
with respect to the SM value, plotted in fig.~6.
This reduction can be at best
about 20{\%}. For comparison we also
show the theoretical prediction on $\bsg /(\bsg )_{SM}$ 
for different values of $\tan \beta$.
The lines in fig.~6 correspond to 
the central values of the input parameters
defined in table 1, and error effects have not been included. 

\section{Conclusions}
\label{sec:conclusions}
We have presented the ${\cal O}(\al )$ QCD corrections to the matching
conditions of the flavour-violating magnetic and chromo-magnetic operators
both in the SM and in  two-Higgs doublet model. We have worked with
an off-shell effective Hamiltonian, and this has allowed us to compute the
relevant Feynman diagrams without the use of asymptotic expansions.

Following this different method we have confirmed the results for the
matching conditions in the SM first found in ref.~\cite{yao}, and recently
recomputed in refs.~\cite{greub,buras2}. Our numerical analysis for the 
branching ratio of
$B\to X_s \gamma$ 
 gives
\beq
\bsg  = (3.62\pm 0.33)\times 10^{-3}  
\eeq
in agreement with previous literature~\cite{misiak1,buras2}.

We have then extended our results to the case of two-Higgs doublet models.
In the case of Model II, the charged-Higgs contribution always increases
the SM prediction. Since the present CLEO measurement lies below the
SM result, $B\to X_s \gamma$ provides quite stringent bounds on the 
charged-Higgs 
mass $M_H$. These bounds are very sensitive to the theoretical errors.
The improvement of the calculation to the NLO significantly
reduces the theoretical uncertainty and therefore tightens considerably
the $M_H$ bounds. Our results are summarised in fig.~5, which
shows the region in the plane $\bsg /\bsl$ versus $M_H$ excluded by our
theoretical calculation. An experimental upper bound on $\bsg /\bsl$ can
then be easily translated into a bound on $M_H$. For instance, using the
present CLEO experimental results and a theoretical uncertainty defined by
one ``Gaussian" standard deviation, we find that charged-Higgs masses below
370 GeV are excluded. This result improves the bound of
about 260 GeV obtained by LO calculations.
We recall that the experimental
results for $\bsl$ show a certain discrepancy between low- and high-energy
measurements (as discussed in sect.~4) and the preliminary ALEPH result
for $B\to X_s \gamma$ is not in perfect agreement with the CLEO measurements.
As shown in fig.~5, developments in the experimental situation can have
an important effect on the $M_H$ bounds.

This bound on $M_H$ does not directly apply to the case of supersymmetric
models, since contributions from other new particles can compensate the
charged Higgs effect, and this is indeed  known to happen in certain 
cases~\cite{io}. Nevertheless, it is clear that this bound is so strong
that it severely restricts the allowed parameter space. 
This has important
consequences on the prediction of new particle masses and on
their observability
at future colliders. 
However, it has to be remarked once again that the impressive
restrictiveness of our
bound on $M_H$ is partly a consequence of the poor agreement between
theory and the CLEO measurement. Developments on the experimental side
or new understanding of non-perturbative effects may then partially relax
this bound.

Finally we have investigated the Model I version of theories with
 two-Higgs doublets.
In this case, the charged-Higgs contribution reduces the SM
value for $\bsg$. Therefore one cannot derive useful bounds on $M_H$,
but instead the theoretical prediction can be brought closer to the CLEO
measurement. However, in view of the limits on $R_b$, the Model I
charged Higgs can reduce the SM prediction for $\bsg$ by at most 20{\%}.

\bigskip

We wish to thank \ F.~Borzumati, \ P.~Chankowski, \ 
K.~Chetyrkin, \ C.~Greub,\   
M.~Mangano,\ 
G.~Martinelli, M.~Misiak, M.~Neubert, M.~Testa, and
C.~Wagner for useful conversations and D.~Schlatter and H.~Wachsmuth for
discussions of the experimental results. We also acknowledge useful
communications
with A.~Buras, G.~Isidori, and N.~Pott.
One of us (G.D.) would like to thank
the Physics Department of New York University for the kind ospitality during
the initial stage of this project.  

\bigskip

Note added: The calculation of the matching condition of $Q_7$ for the
two-Higgs doublet model II, in the large $\tan \beta$ limit,
has also been recently presented in ref.~\cite{strum}, and the result
agrees with our eq.~(\ref{higceq}). In the final version of this paper,
various formulae entering the calculation of $BR(B\to X_s\gamma)$ have been
corrected, following suggestions by F.~Borzumati, C.~Greub, M.~Misiak and
M.~Neubert.


\begin{thebibliography}{99}
\bibitem{cleoks} R.~Ammar {\it et al.} (CLEO Collab.), 
\prl{71}{93}{674}.
\bibitem{cleo} M.S.~Alam {\it et al.} (CLEO Collab.), \prl{74}{95}{2885}.
\bibitem{sav} A.F. Falk, M. Luke, and M. Savage, \pr{49}{94}{3367}.
\bibitem{tran} N.G. Deshpande, X.-G. He, and J. Trampetic, \pl {367}{96}{362};\\
J.M. Soares, \pr{53}{96}{241};\\
G. Eilam, A. Ioanissian, R.R. Mendel, and P. Singer, 
{\it Phys. Rev.} {\bf D53} (1996) 3629.
\bibitem{vol} M.B. Voloshin, \pl{397}{97}{295}.
\bibitem{isi} G. Buchalla, G. Isidori, and S.J. Rey, preprint hep-ph/9705253.
\bibitem{mcc} A. Khodjamirian, R. R\"uckl, G. Stoll, and D. Wyler, 
{\it Phys. Lett.} {\bf B402} (1997) 167;\\
Z. Ligeti, L. Randall, and M.B. Wise, \pl{402}{97}{178};\\
A.K. Grant, A.G. Morgan,  S. Nussinov, and  R.D. Peccei,
{\it Phys. Rev.} {\bf D56} (1997) 3151.
\bibitem{buras1} A.J.~Buras, M.~Misiak, M.~M\"unz and S.~Pokorski, \np
{424} {94} {374}.
\bibitem{ciuchini1} M.~Ciuchini {\it et al.}, \pl{334}{94}{137}.
\bibitem{yao} K.~Adel and Y.P.~Yao, \pr{49}{94}{4945}.
\bibitem{ali} A.~Ali and C.~Greub, \zp{49}{91}{431}; \pl {259}
{91} {182}; \zp{60}{93}{433};
\pl {361} {95} {146}.
\bibitem{wyler} C.~Greub, T.~Hurth and D.~Wyler, \pl{380}{96}{385}; 
\pr {54} {96} {3350}.
\bibitem{misiak1} K.~Chetyrkin, M.~Misiak and M.~M{\"u}nz, \pl
{400} {97} {206}.
\bibitem{buras2} A.J.~Buras, A.~Kwiatkowski and N.~Pott, preprint
TUM-HEP-287-97 [hep-ph/9707482].
\bibitem{aleph} F.~Parodi, talk presented at the 32nd Rencontres de Moriond,
March 1997. 
\bibitem{greub} C.~Greub and T.~Hurth, Phys.~Rev. D56 (1997) 2934.
\bibitem{smirnov} See V.A.~Smirnov, \mpl{A10}{95}{1485} and references
                  therein.
\bibitem{curci} G.~Cella, G~.Curci, G.~Ricciardi and A.~Vicer{\'e}, \np
{431} {94} {417}.
\bibitem{simma} H.~Simma, \zp{61} {94}{67}.
\bibitem{wilson} K.~Wilson, \pr{179}{69}{1499}.
\bibitem{witten} E.~Witten, \np{122}{77}{109}.
\bibitem{abbott} L.F.~Abbott, \np{185}{81}{189}.
\bibitem{collins} J.C.~Collins, {\it Renormalization}, Cambridge University
Press (1984).
\bibitem{fuj} K.~Fujikawa, \pr{7}{73}{393}.
\bibitem{deshpande} N.G.~Deshpande and M.~Nazerimonfared, \np{213}
{83}{390}.
\bibitem{altarelli} G.~Altarelli and L.~Maiani, \pl{52} {74} {351};\\
M.K.~Gaillard and B.W.~Lee, \prl{33} {74} {108}.
\bibitem{misiak2} M.~Misiak, \pl{269} {91} {161}.
\bibitem{ciuchini2} M.~Ciuchini {\it et al.}, \pl {316}{93}{127};\\
M.~Ciuchini, E.~Franco, L.~Reina and, L.~Silvestrini, {\it Nucl. Phys.}
{\bf B421} (1994) 41.
\bibitem{gm2} See for example A.~Czarnecki and B.~Krause,
         {\it Nucl.~Phys.~Proc.~Suppl.}~{\bf 51C}, (1996), 148. 
\bibitem{dav} A.T.~Davydychev and B.~Tausk, \np{397}{93}{123}.
\bibitem{SW} H.~Simma and D.~Wyler, \np{344}{90}{283}.
\bibitem{cabib} N. Cabibbo and L. Maiani, \pl{79}{78}{109};\\
Y. Nir, \pl{221}{89}{184}.
\bibitem{neubr} M. Neubert, Phys. Rep. 245 (1994) 259;\\
Int. J. Mod. Phys. A11 (1996) 4173.
\bibitem{pott} N.~Pott, \pr {54} {96} {938}.
\bibitem{borz} F.~Borzumati and C.~Greub, preprint ZU-TH 31/97 [hep-ph/9802391].
\bibitem{pdg} Particle Data Group, R.M. Barnett {\it et al.}, \pr{54}{96}{1}.
\bibitem{cdf} S. Leone, for the CDF Coll., Proc. 
 High-Energy Physics International Euroconference
on Quantum Chromodynamics: QCD 97, Montpellier, France, 3-9 Jul
1997. 
\bibitem{d0} K. Genser, for the D0 Coll., Proc.
11th Les Rencontres de Physique de la Vallee d'Aoste: Results and
Perspectives in Particle Physics, La Thuile, Italy, 2-8 Mar 1997.
\bibitem{neub2} M. Neubert, in Heavy Flavours, ed. by A.J. Buras and
M. Lindner, World Scientific, Singapore, preprint  hep-ph/9702375.
\bibitem{richman} J.D. Richman, Proc. 28th International Conference 
on High-energy Physics (ICHEP 96), Warsaw, Poland, 25-31 Jul
1996. 
\bibitem{cleosl} B. Barish {\it et al.} (CLEO Coll.), \prl{76}{96}{1570}.
\bibitem{argussl} H. Albrecht {\it et al.} (ARGUS Coll.), 
\pl{318}{93}{397}.
\bibitem{lepwg} LEP Collaborations, LEP Electroweak Working Group,
SLD Heavy Flavour Group, preprint LEPEWWG/97-02.
\bibitem{feig} M. Feindt, talk at International Europhysics Conference
on High-Energy Physics, Jerusalem, Israel, 19-26 August 1997.
\bibitem{nat} S.L. Glashow and S. Weinberg, \pr {15} {77}{1958}.
\bibitem{chwil} W.S. Hou and R.S. Willey, \pl {202}{88}{591};\\
B. Grinstein, R. Springer, and M. Wise, \np{339}{90}{269}.
\bibitem{higlo} J.L. Hewett and J.D. Wells, \pr{55}{97}{5549};\\
 T. Hayashi, M. Matsuda, and M.
Tanimoto, \ptp{89}{93}{1047};\\
J.L. Hewett, \prl{70}{93}{1045};\\
V. Barger, M. Berger, and R.J.N. Phillips, \prl{70}{93}{1368};\\
S. Bertolini, F. Borzumati, A. Masiero, and G. Ridolfi, {\it Nucl. Phys.}
{\bf B353} (1991) 591.
\bibitem{pok} M. Misiak, S. Pokorski, and J. Rosiek, in
``Heavy Flavors II", eds. A.J. Buras and M. Lindner, World Scientific,
Singapore, hep-ph/9703442;\\
P. H. Chankowski and S. Pokorski, in ``Perspectives in supersymmetry",
ed. G.L. Kane,
World Scientific, Singapore, hep-ph/9707497.
\bibitem{anl} H. Anlauf, \np{430}{94}{245}.
\bibitem{io} R. Barbieri and G.F. Giudice, \pl{309}{93}{86}.
\bibitem{lim} V. Barger, J.L. Hewett, and R.J.N. Phillips, \pr{41}{90}{3421};\\
G. Buchalla {\it et al.}, \np{355}{91}{305};\\
Y. Grossman, \np{426}{94}{355}.
\bibitem{strum} P. Ciafaloni, A. Romanino, and A. Strumia, preprint
hep-ph/9710312.
\end{thebibliography}
\end{document}